\documentclass[12pt,english,preprint]{aastex}
\usepackage[T1]{fontenc}
\usepackage[latin9]{inputenc}
\usepackage{textcomp}
\usepackage{amstext}
\usepackage{graphicx}

\makeatletter

\providecommand{\tabularnewline}{\\}

\newcommand{\degreee}{$^\circ$}

\newcommand{\sbright}{B$_\odot$}

\shorttitle{Solar Wind Observations with \emph{STEREO}/HI-2}
\shortauthors{DeForest et~al.}

\makeatother

\usepackage{babel}

\begin{document}

\title{Observations of Detailed Structure in the Solar Wind at 1 AU with
\emph{STEREO}/HI-2 }

\author{C.~E. DeForest}

\affil{Department of Space Studies, Southwest Research Institute, Boulder,
CO 80302}

\email{deforest@boulder.swri.edu}

\author{T.~A. Howard}

\affil{Department of Space Studies, Southwest Research Institute, Boulder,
CO 80302}

\email{howard@boulder.swri.edu}

\author{S.~J. Tappin}

\affil{National Solar Observatory, Sunspot, NM 88349}

\email{jtappin@nso.edu}
\begin{abstract}

We present images of solar wind electron density structures at
distances of 1 A.U., extracted from the \emph{STEREO}/HI-2 data.  Collecting
the images requires separating the Thomson-scattered signal from the
other background/foreground sources that are $10^3$ times brighter.
Using a combination of techniques, we are able to generate calibrated
imaging data of the solar wind with sensitivity of a few
$\times10^{-17}$\sbright, compared to the background signal of a few
$\times10^{-13}$\sbright, using only the \emph{STEREO}/HI-2 Level 1 data as
input.  These images reveal detailed spatial structure in CMEs and the
solar wind at projected solar distances in excess of 1 AU, at the
instrumental motion-blur resolution limit of 1\degreee{} -
3\degreee. CME features visible in the newly reprocessed data from
December 2008 include leading-edge pileup, interior voids, filamentary
structure, and rear cusps. ``Quiet'' solar wind features include V
shaped structures centered on the heliospheric current sheet,
plasmoids, and ``puffs'' that correspond to the density fluctuations
observed in-situ. We compare many of these structures with in-situ
features detected near 1 AU. The reprocessed data demonstrate that it
is possible to perform detailed structural analyses of heliospheric
features with visible light imagery, at distances from the Sun of at
least 1 AU.
\end{abstract}

\keywords{Sun: coronal mass ejections (CMEs), solar-terrestrial relations---Solar
System: interplanetary medium---methods: data analysis}

\section{Introduction}

The last decade has seen the emergence of a new class of white light
imaging instruments, heliospheric imagers, that combine precise
photometric wide-field imaging with extremely deep baffles.  Reviews
of the capabilities of heliospheric imagers have been written by
\citet{Harrison2011} and by \citet{Howard2011}.

Heliospheric imagers are in principle capable of detecting and
tracking both large- and small-scale transients in the solar wind by collecting and
imaging sunlight that is scattered directly off of free electrons in
interplanetary space, just as coronagraphs capture the K-corona in the
vicinity of the Sun.  The first proof-of-concept observations of
heliospheric transients were demonstrated with the Helios zodiacal
light instrument \citep{Richter1982,Jackson1985} and the first
wide-field dedicated heliospheric imaging instruments were the Solar
Mass Ejection Imager (SMEI), launched in 2003 on the \emph{Coriolis} spacecraft \citep{Eyles2003} and
the \emph{STEREO} mission Heliospheric Imagers (HIs), launched in 2006
\citep{Eyles2009}. These instruments have proven to be effective
for observing solar wind transient phenomena such as coronal mass
ejections \citep[e.g.][]{Howard2006,Harrison2008,Webb2009}, corotating
interaction regions \citep[e.g.][]{Sheeley2008,Rouillard2008,Tappin2009}, and solar wind
``puffs'' \citep[e.g.][]{Rouillard2010} and ``blobs''
\citep[e.g.][]{Sheeley2009,Sheeley2010}. 

Small, non-CME structure (puffs and blobs) have been observed in the
solar wind in the outer corona \citep[e.g.][]{Sheeley1997}, by in-situ
spacecraft \citep[e.g.][]{Schwenn1990,Phillips1995}, using
interplanetary scintillation \citep[e.g.][]{Rickett1991}, and by the
\emph{STEREO}/HI-1 instruments \citep[e.g.][]{Clover2010}.  Quantitative, detailed
imaging of these structures has remained elusive at elongation angles above
40\degreee (in the \emph{STEREO}/HI-2 field of view).

The greatest challenge in heliospheric imaging is signal extraction of
the faint Thomson scattering signal from the far brighter foreground
and background signals. Large CMEs have intensities (at 45\degreee\ 
solar elongation) of order $10^{-14}$ \sbright, and the faintest
remotely tracked solar wind features (puffs) have intensities of the
order of a few $\times\ 10^{-16}$
\sbright\ \citep{Rouillard2010,Tappin2009}.  (1 \sbright\ is a solar
brightness unit $\sim$ the surface brightness of the solar photosphere).
\citet{Tappin2009} have demonstrated that careful analysis of
differenced elongation-vs-time images (``J-maps'' or ``J-plots'') can detect
features with brightnesses of a few $\times 10^{-16}$ \sbright, and
\citet{Rouillard2010} has performed simlar J-plot analyses of small
transient structures propagating with the ambient wind and becoming
entrained in a CIR.  Both groups used the structure of the J-plot
imagery to reveal morphology of features with signal strength at or
below the noise level imposed by the background star field in their
processed data.  To our knowledge no group has yet produced clear 2-D
imagery of evolving wind features in HI-2 at comparable or greater
sensitivity.

These faint signals are detected by \emph{STEREO}/HI against a background
dominated by three principal sources: instrumental stray light, which
forms a fixed pattern on the detector; zodiacal light (which we also
refer to as ``F corona'') at the few $\times 10^{-13}$ \sbright\ level
that varies only slowly as the spacecraft orbits the Sun; the
background starfield, which is extremely spiky ($10^{-14}$
\sbright\ for a 10th magnitude star in the HI-2A instrument) and
drifts with the orbital motion of the spacecraft
\citep{Eyles2003}; and small artifacts, reminiscent of cosmic rays or
dust, that are visible in many images despite on-board processing that
removes most cosmic ray impacts. SMEI is also subjected to saturation
from magnetospheric particles, auroral glow, and moonlight. The stray
light pattern in \emph{STEREO}/HI is weak compared to the F corona.

In the present paper, we describe and demonstrate a technique by which
the HI-2 data can be background-subtracted to reveal solar wind
structures throughout the field with residual noise levels of $\sim
3\times 10^{-17}$ \sbright\ in 1\degreee\ square patches of image, within
a factor of three of the photon counting noise level (which is
$\sim 10^{-17}$ \sbright\ against a background of $10^{-13}$
\sbright\ in the standard HI-2 data set, when averaged over a 1\degreee\ 
square patch).

We have reprocessed HI-2A data across a time period of several days
surrounding a well-documented CME that occurred in December 2008.  The
reprocessed data have a noise floor lower than has been available in
previous analyses, and reveal rich detail in the solar wind and CME
structure, that has been previously inaccessible.

Along with the CME we identify several solar wind transient
features, including the CME itself, a forerunner, a filament, and some
small transient structures that may correspond to ``puffs'' and
``blobs'' observed by \citet{Rouillard2010} and \citet{Sheeley2009}
respectively.   

Because the technique does not require difference subtraction, we are
also able to identify cavity regions surrounding the CME event
itself that have been undetectable with previous analytical
techniques.  Making J-plots from the resulting cleaned movie shows
striking correspondence between the apparent arrival times of the
observed heliospheric features at Earth, and the detection of similar
density structures by the \emph{Wind} spacecraft.

We conclude with discussions of the limitations of image processing
for extracting solar wind heliospheric image data, of physical
implications of high quality imaging of transients in the solar wind, 
and planned work to exploit the newly available data sets.

\section{Instrument and Event}

We consider the outer heliospheric imager (HI-2A) on board the
\emph{STEREO-A} (``ahead'') spacecraft
\citep{Eyles2009,RHoward2008,Kaiser2008}. HI consists of two cameras,
HI-1 normally observes across an elongation range of 4\degreee\ to
24\degreee\ and HI-2 from 19\degreee\ to 89\degreee, both centered on
the line formed by the image-projected ecliptic plane. \emph{STEREO} was
launched so that its angular separation from the Sun-Earth line
increased at a rate of $\sim$22.5\degreee\ per year.  During the time
period of interest (December 2008), \emph{STEREO-A} was 42\degreee\ away from
the Sun-Earth line and at a radial distance of 0.97 AU from the Sun.
HI-2 operates on a 2 hour cadence, summing many individual camera
frames into 7,000 second ``macro-exposures'' on board the spacecraft.
The macro-exposures are despiked on-board in the temporal direction to
remove cosmic rays and similar artifacts before summing.  The camera
has an active image planes of $2048\times 2048$ pixels but is summed
$2\times 2$ before downlink, so that reduced images are $1024\times
1024$ pixels in size.

We chose to focus on HI-2A alone rather than a joint study with HI-2A
and HI-2B (on the ``behind'' spacecraft) because the starfield in
HI-2B images is more difficult to isolate than from HI-2A, due to an 
apparent defocus seen in the HI-2B images.  That defocus is smaller than
the expected motion blur we describe (1\degreee - 3\degreee depending on 
feature speed), but large enough to require slightly different techniques
for starfield subtraction than are described here.

The event of primary interest was a CME that erupted from the Sun on
12 December 2008 and reached the HI-2 field of view the following
day. It is generally regarded as the first geoeffective CME observed
during the \emph{STEREO} mission and has been documented by
\citet{Davis2009}, \citet{Liu2010} and \citet{Lugaz2010}.  Our
observational coverage includes the time period from 11--21 December,
which includes the pre- and post-CME solar wind in this region.  We
set out to extract and observe whatever structure we could from the
event as it propagated outward.

\section{Processing}

We removed background from the HI-2 images in stages.  Like other
unpolarized coronagraphs such as \emph{SOHO}/LASCO, the HI-2 images include a
strong image component from dust (the zodiacal light or ``F corona''),
which cannot in principle be separated from the fixed, smooth bulk
component of the desired Thomson-scattered (``K-corona'') signal.
Because the F corona is nearly constant, it can be separated from
dynamic transients in the K coronal signal by tracking and subtracting
the minimum brightness in any single pixel on the focal plane over an
interval long compared to the crossing time of the features of
interest.  This yields images of the ``excess density'' due to
transient features. A major complication for HI-2 is the background
starfield. Unlike coronagraph data, both the F and K coronal signals
in the HI-2 field are faint compared to the bright stars in the starfield.
Further, the starfield has different motion characteristics than the F
corona because of spacecraft orbital motion, and is extremely spiky,
stretching the limits of pattern identification and rectification.

Because the Thomson signal is so faint, additional steps are required
to remove second-order effects not visible in less challenging data,
which requires two additional filtration steps.  Noise sources are
summarized in Table \ref{tab:noise-sources}. Two bright planets were
visible in the field of view -- Earth and Venus -- and they saturated
the detector, leaving top-to-bottom saturated streaks that damaged
some of the later filtering steps.  We cropped the field of view of
our observation to eliminate them.

\begin{figure}
\center{\includegraphics[width=3in]{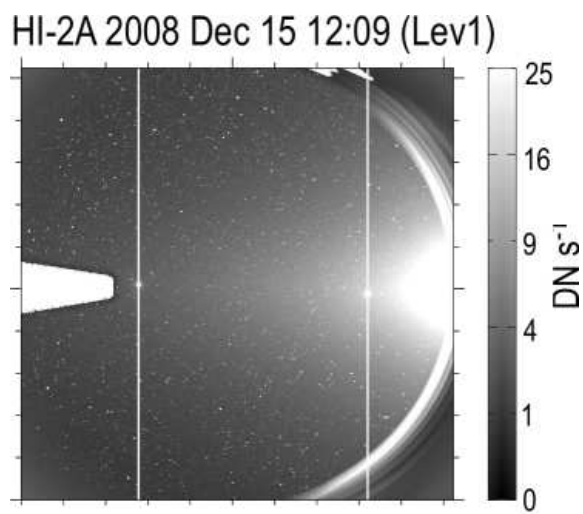}}
\caption{\label{fig:raw-data}Unprocessed \emph{STEREO}/HI-2 Level 1 image at
  0-25 DN s$^{-1}$ (square root scaled to increase dynamic range), obtained
  on 15 December 2008 at 12:09 UT (2 hour accumulation). The Sun is to
  the right, but well outside the image, and the coordinates are in
  units of image pixels. Dominant features include the F corona,
  starfield, and CCD artifacts including the two planets Earth (left)
  and Venus (right). The planets saturate the CCD causing ``bleeding''
  of charge in the vertical direction and defining the width of the
  usable field of view for our analysis. The K coronal signal of interest
  has an amplitude of ~0.02 DN s$^{-1}$, 1000 times fainter.
}
\end{figure}

We began our analysis with \emph{STEREO}-supplied Level 1 data, which are
linearly calibrated (no nonlinear calibration applied) in normalized
digitizer counts per second (DN s$^{-1}$). An example frame is given in
Figure \ref{fig:raw-data}, which has been square-root scaled to show
the dominant features in the data: bright stars interspersed with the
F corona at exposure levels of a few tens of DN s$^{-1}$.  The
processing includes five major steps: stationary background removal;
celestial background removal (including cross-image distortion
measurement); residual F corona removal; moving-feature filtration in
the Fourier plane; and conversion back to focal plane coordinates. The 
K coronal signal is not visible at this brightness scale, but is of order 
0.02 DN s$^{-1}$ compared to the scale of 25 DN s$^{-1}$.

\begin{table}
\caption{\label{tab:noise-sources}Characteristics of background and noise
sources in the \emph{STEREO}/HI-2 data}

\centering{}\begin{tabular}{|c|c|c|}
\hline 
Background source & Motion & Amplitude (DN s$^{-1}$)\tabularnewline
\hline
\hline 
Stray light & stationary & up to 30 at bright edge\tabularnewline
\hline 
F corona  & approx. stationary & up to 30\tabularnewline
\hline 
Starfield & moves 1$^{\circ}$/day & up to 20\tabularnewline
\hline 
Galaxy & moves 1$^{\circ}$/day & up to 20\tabularnewline
\hline
\end{tabular}%
\end{table}

\subsection{\label{sub:Stationary-background-removal}Stationary background removal}

\begin{figure}
\center{\includegraphics[width=6in]{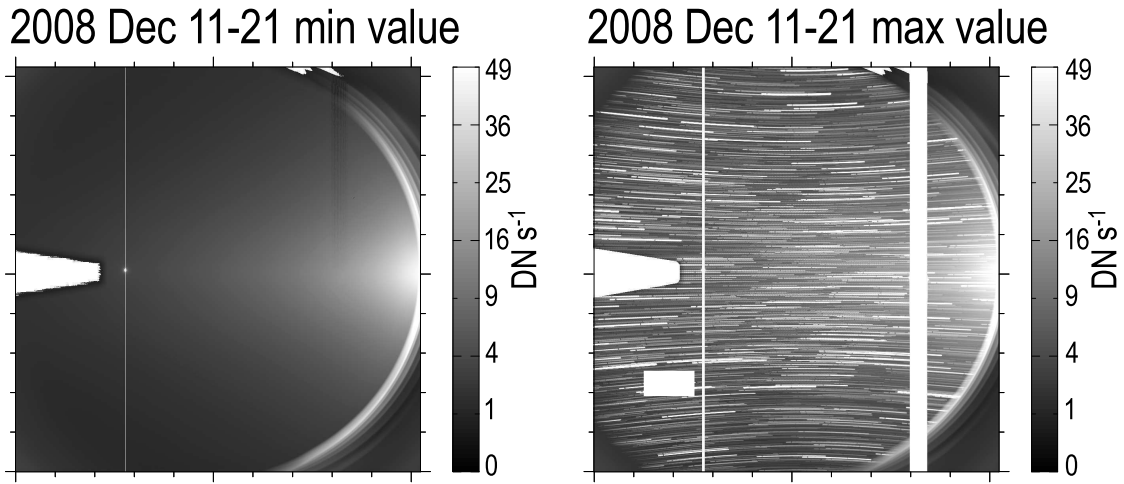}}

\caption{\label{fig:Percentile-images}Faintest- and brightest-value
  images from the \emph{STEREO}/HI-2A Level 1 data spanning 2008 Dec 11-21.
  (Left) Faintest-value image shows F corona and stray light; (Right)
  brightest-value image shows moving starfield, spikes from
  near-spacecraft dust or cosmic rays, and marked data dropouts.
  The faint image was further processed to reduce starfield, and kept
  as the F corona model for the data set. Both images have been scaled
  to 0-50 normalized HI-2 DN $s^{-1}$ and square root scaled to display faint
  features.}
\end{figure}

The F corona and stray light signals are both approximately stationary
on the image plane.  We do not attempt to separate them, but remove
both with the same image processing step.  We identified the
stationary background by simple stacking of the complete 11 day data
set.  We treated each pixel as a statistical population of
brightnesses, and examined the faintest brightness value from each
pixel.  Data dropouts are set by the \emph{STEREO} pipeline to have negative
numbers. For the sorting, we treated these numbers as large, so that
the faintest valid values of each pixel could be considered. The
faintest and brightest value images for the data set spanning 11-21
December 2008 in HI-2a are shown in Figure
\ref{fig:Percentile-images}. The faintest-value image contains
primarily F corona and stray light. The brightest-value image on the
right of Figure \ref{fig:Percentile-images} clearly shows the
starfield as a set of streaks: as each bright star passes across the
field of view over the data set, it sets the brightest value of that
pixel. The optics distortion function is visible as a variation in the
characteristic length and angle of the streaks across the field.

\begin{figure}
\center{\includegraphics[width=3in]{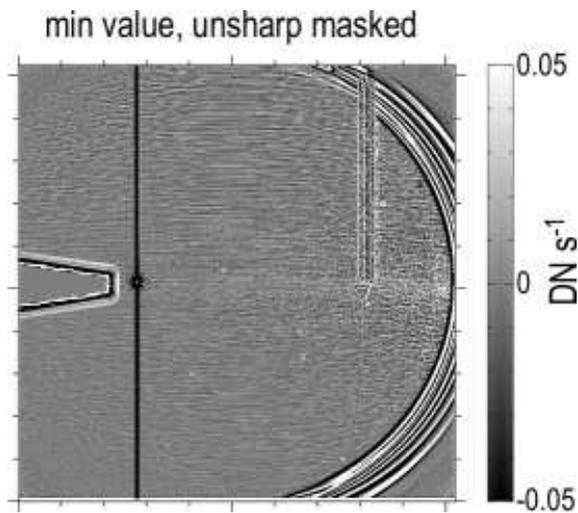}}

\caption{\label{fig:minimum-value image}Minimum value image from Fig. \ref{fig:Percentile-images}
unsharp masked with a 9$\times$9 pixel boxcar to show detail. Residual streaks
from the starfield are visible throughout, as are artifacts from planetary
overexposure and speckles from statistical variation in the F corona.}
\end{figure}

Measuring the F corona and stray light requires slightly more care
than simply extracting the minimum value. Three important artifacts
remain: overexposure streaks from planets and other bright features;
residual streaking from the starfield; and statistical noise-induced
errors from the F corona. The first two can be combatted with in-image
filtering; the latter is more problematic. Figure \ref{fig:minimum-value image}
shows the minimum value image, unsharp-masked to highlight these remaining
structures that are present at the \textpm{}0.1 DN s$^{-1}$ level.
The exposure artifacts can be removed by using a low-percentile image
(we chose the 5th percentile in pixel brightness) at some cost in
additional starfield artifacts. The starfield can be further reduced
with spatial median filtration, and we use a 5$\times$5 median filter on
our fixed background image.  This technique is similar to that used
used by the \emph{STEREO}/HI-2 official pipeline, which involves taking a per-pixel
average of the lowest quartile of values in an 11 day window.

The speckling toward the right in Figure \ref{fig:minimum-value image}
is more problematic than the other two error sources in the fixed
pattern image, and we attribute it to variations in the noise level of
different detector bins on the CCD, coupled with the skewed sample
imposed by a minimum-value measurements. The Level 1 data are linearly
flat-field corrected and blur compensated for exposure during readout,
but it is of course not possible to eliminate all detection noise, and
the data value at each pixel is the sum of the image plus samples of
independent random variables from photon counting and the inherent
fixed-pattern noise of the CCD. The minimum-value operation across an
ensemble of raw images necessarily produces a skewed sample of these
detection noise sources. The two sources introduce two visually
distinct signatures.  The first is due to the fixed pattern noise of
the CCD and appears as ``speckle'' in the brightest portion of the
image (center-right).  The second is due to skewed sampling of the
photon noise from the F corona itself. It yields both a
brightness-dependent level of ``snow'' in Figure
\ref{fig:minimum-value image} and a systematic underestimation of the
background brightness. It may in principle be possible to correct for
this underestimation, by characterizing the independent statistics of
the fixed pattern noise and photon counting noise.

\begin{figure}
\center{\includegraphics[width=3in]{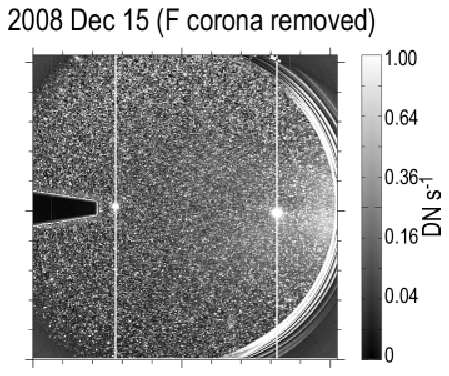}}

\caption{\label{fig:F subtraction}Background-subtracted version of Figure
\ref{fig:raw-data}, using the background derived via the 5th percentile
image of an 11 day dataset, with skewed-median smoothing as described
in the text. The dynamic range is 25$\times$ more sensitive than Figure \ref{fig:raw-data}.
The fixed brightness pattern containing the F corona has been reduced
by a factor of approximately 300 but is still faintly visible at right.  
Many individual stars greatly exceed the dynamic range of this plot.}
\end{figure}

A nice compromise between the rabbit hole of independent statistical
treatment of noise random variables in each pixel, and simply ignoring
the problem, is skewed median filtering of the background. Selecting
the 30th percentile value from each 5$\times$5 pixel neighborhood
removes much of the local brightness from the faint starfield at the
cost of increasing F coronal leakage slightly from the skewed noise
sample.  Figure \ref{fig:F subtraction} shows the quality of the fit:
the non-stellar fixed pattern is reduced by a factor of approximately
300 by background subtraction with the skewed-median background
image. As with other broadband unpolarized Thomson scattering data
\citep[e.g.\ from LASCO C3 ][]{Morrill2006} it is not possible to
separate the F coronal (zodiacal light) signal from the fixed portion
of the K coronal (Thomson scattered) signal. The subtracted image also
demonstrates why there is stellar structure in the minimum image. At
the faint end of the stellar brightness distribution, there are stars
in most pixels with only isolated darker ones.

\subsection{\label{sub:Celestial-background-removal}Celestial background removal}

Removing the stationary and quasi stationary background components
leaves the celestial background of stars, galaxies, and other features
that remain fixed on the celestial sphere rather than the image plane.
The starfield is the most challenging noise source to remove, because
individual stars take the forms of spikes on the image, with high
gradients. Subtraction of a model starfield requires deeply sub-pixel
alignment on the image. The alignment is made more challenging by
the fact that the starfield moves on the image plane between exposures
as the spacecraft orbits the Sun, and by the fact that the optics
impose a nontrivial projection function on the wide-angle field of
view. We generated a model starfield from the data themselves, fitting
the distortion function $D$ between images as the composition of
a parameterized deprojection function from the image plane to the
celestial sphere. This is followed by a time-dependent orbital transformation,
and reprojection onto the focal plane: \begin{equation}
D_{q_{i}}(x,y,\Delta t)=\left(P_{q_{i}}\circ S_{q_{i}}(\Delta t)\circ P_{q_{i}}^{-1}\right)(x,y)\label{eq:distortion}\end{equation}
 where $P_{q_{i}}$ is the projection function from celestial coordinates
to image coordinates and is invariant under nominal operations (though
it includes a rotation to orbital coordinates that could in principle
change if the instrument pointing changed from the nominal ecliptic-plane
orientation), $S_{q_{i}}$ is the orbital-shift function in ecliptic
coordinates and contains all the time dependence in $D$, and the
$q_{i}$ are the fit parameters. As a check
on the projection correctness we fitted the orbital speed and spacecraft attitude
as part of the parameterization. The fit was
accomplished by measuring the in-plane coordinate shift between the
starfields in a collection of points from two separate images collected
five days apart, and adjusting both $P$ and $S(2.5d)$ to minimize
the RMS error between the modeled and measured shifts. Because $S$
is approximately linear in time (treating the orbit as circular), 
and $P$ is invariant across exposures, this yielded
a coordinate transformation capable of transforming the starfield
between the focal planes of any two exposures at known times in the
same general part of the \emph{STEREO} orbit.

\subsection{\label{sub:cross-image-distortion}Measuring cross-image distortion}
Fitting the functions in Equation (\ref{eq:distortion}) requires measuring
$D(x,y)$. We measured the function with a regular grid of tiepoints
that were used as the basis of correlative fits between two images
taken 5 days apart. In our first attempt, we centroided bright stars (on the
irregular grid formed by the stars themselves); but the centroid identifications
had positional noise at about the 0.2 pixel RMS level, with occasional
outliers (in bright stars) of about 0.5 pixel displacement.
We attribute this noise source to subpixel variations in detector sensitivity,
coupled to the strong gradients in the stellar images, as described
by \citet{Jackson2004}. Switching to patch correlation
between corresponding patches in the earlier and later images yielded
much lower noise, which we attribute to dithering of the subpixel
gradients across all the stars in each patch (typically dozens to
hundreds of faint stars), thereby beating down this noise source via multiple
sampling.  A further noise source, which is important even to the correlative
fits near the edges of the image, is variation in the instrument PSF 
across the image field of view.  Near the edges of the field of view, 
stars become visible elongated and the correlative fits appear correspondingly
noisier.  PSF effects are not important near the center of the image plane.

We arrived at a square correlated patch of 31$\times$31 pixels in size, large 
enough to yield a large central correlation spike at the proper alignment 
and null the correlation signal for spurious alignments of different bright
stars between the two images. Typical patches contained dozens of
individually resolved stars in addition to the unresolved faint starfield.
To better equalize the correlation contribution between bright and
faint stars, we took the square root of each image patch. Typical
correlation coefficients with 1 pixel offset were 0.4--0.6, though
a few coincidental missed fits yielded local maximum correlation coefficients
as high as 0.80 for misaligned starfields. For the proper subpixel
displacement typical correlation coefficients were above 0.995. The
patches were chosen on a regular grid, and missed fits were rejected.
In general, fits near the edge of the field of view were missed, along
with fits that included a planetary saturation spike in one or the
other image.

\begin{figure}

\begin{centering}
\center{\includegraphics[width=6in]{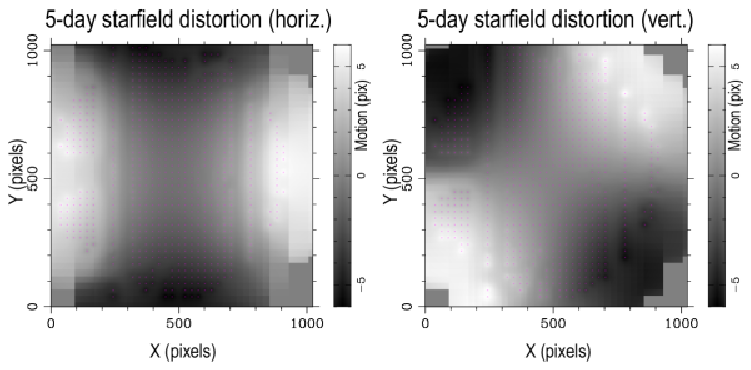}}
\par\end{centering}

\caption{\label{fig:measured-distortion}Initial measured nonlinear
  distortion function for \emph{STEREO}/HI-2A shows measured horizontal and
  vertical distortions as a function of position over a 5 day interval
  in Dec 2008 in the HI-2A field of view. Kept correlation patch
  centers are marked and form the data comparison points for the
  subsequent fit of the coordinate transformation.  For display, the
  images are interpolated between patch centers using an inverse
  square distance-weighted average.}

\end{figure}

The initial guess offset, which we used to find the corresponding
patch on the later image and seed the correlation fit, was formed
based on a horizontal offset of 72 pixels in 5 days. We calculated the
correlation coefficient between the two images 5 days apart on a
15$\times$15 grid of integer pixel offsets around the horizontal
offset location, and the subpixel fit was seeded with the best offset
among the trial offsets. Further alignment was fit using a
simplex/amoeba fitter \citep{PressBook} with initial simplex radius of
0.75 pixels, centered on the best trial alignment. Patches were
rejected if either the original or the shifted patch, with margins,
intersected the edge of the telescope image plane or contained even one
saturated pixel, or if the final best-fit correlation coefficient
between the patches in the earlier and later frame was below
0.99. Figure \ref{fig:measured-distortion} shows the horizontal and
vertical components of the measured $D(x,y,2.5d)$, after removal of
the base horizontal 72-pixel offset.

The $D_{qi}$ distortion function was assembled from parameterized
components as in Equation (\ref{eq:distortion}). The amount of displacement
in $S$ was a parameter of the fit, to verify that the spacecraft's
orbital speed was reproduced by the fitter. The $P_{qi}$ was a perspective
projection of the celestial sphere with variable camera angle and 
$\mu$ parameter (distance from projecting point to the center of
the sphere), as described by \citet{Snyder1987} and \citet{Brown2009}. We
hoped to use the starfield, as did Brown et~al., but to characterize 
the frame-to-frame offsets to better than their reported 1 pixel precision
by refining the family of coordinate transformations used to relate
celestial and \emph{STEREO} focal-plane coordinates.

The initial guess $P$ was a stereographic projection of the celestial
sphere (i.e. $\mu=1$) and the initial guess $S$ was based on a
1\degreee{}~d$^{-1}$ rotation about an axis parallel to the vertical
axis of the image. The $q_{i}$ were iterated through an amoeba fitting
process to find the best overall $D$ using the sum-of-squares distance
between the two sets of tiepoint pixel coordinates after
transformation to the central time between the two images.

Initial free parameters were: pointing of the projection center
relative to image center (X and Y), roll and ``B'' (out-of-plane)
angle of the projective plane relative to the ecliptic pole, orbital
rotation $\theta$, and projection distortion parameter $\mu$. The roll
and B angles, in particular, were treated as instrumental constants
across the dataset. This is a valid simplification for a single
few-day data set, though for a more general pipeline process the two
angles should be allowed to change according to spacecraft orbital
phase relative to the celestial nodes of the orbit.

The initial fitting process led to a 6-parameter coordinate
transformation with a residual RMS error of 0.35 pixel between the
measured and modeled distortion parameters at the kept tiepoints,
compared to the initial error (depicted in Figure
\ref{fig:measured-distortion}) of 3.6 pixels.  An order of magnitude
improvement isn't bad, but more improvement was possible. The
residuals contained obvious errors in a quadrant pattern for vertical
displacement and in a vertical striation pattern for horizontal
displacement, and each was treated with ad-hoc perturbation functions
of known pattern and variable amplitude, adding two more parameters to
the fit. The horizontal displacement ad hoc distortion
was \begin{equation} x'=x+\beta
  \sin\left(\frac{3\pi}{2}\frac{(x-512)}{512}\right)\cos\left(\frac{\pi}{2}\frac{(y-512)}{512}\right)\label{eq:x-distortion}\label{eq:adhoc1}\end{equation}
and the vertical displacement ad hoc distortion was \begin{equation}
  y'=y+\alpha
  \sin^{2}\left(\frac{2\pi}{3}\frac{(x-512)}{512}\right)\sin\left(\frac{\pi}{2}\frac{(y-512)}{512}\right)\label{eq:y-distortion}\label{eq:adhoc2}\end{equation}
where $\alpha$ and $\beta$ are the fitted amplitude coefficients and
$x$ and $y$ are pixel index coordinates in the original image plane
before any other coordinate transforms were applied,
i.e. \begin{equation}P_{final}^{-1}=P_{6-param}^{-1}\circ f_{adhoc}(x,y).\label{eq:adhoc-application}\end{equation}  
With the
$\alpha$ and $\beta$ parameters, a total of eight fit parameters were
fitted to the measured displacements of each patch.  The ad hoc
functions had the correct performance for the central region of the
image plane, but diverged slightly in the periphery of the valid
images. For the purpose of this exploratory study, we simply discarded
the region of poorer convergence, limiting the study to a 400$\times$750
pixel region at the center of the image plane. Limiting the field of
view is a good choice as it limits other edge effects such as
vignetting and PSF variation at the periphery of the image plane, and
also allows us to avoid the two top-to-bottom saturated regions in
this data set (one from Earth and one from Venus).

\begin{figure}
\begin{centering}
\center{\includegraphics[width=6in]{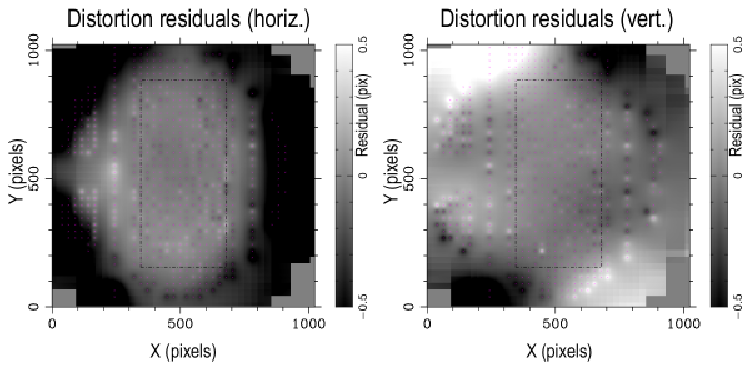}}
\par\end{centering}

\caption{\label{fig:Residual-distortion-error}Residual distortion error after
starfield distortion fitting is 0.09 pixel RMS in the central rectangle
(compare to Figure \ref{fig:measured-distortion}). Locations where the distortion
was measured are marked.}
\end{figure}

After iterating the distortion fit with the two ad-hoc correction functions,
the distortion residual error was 0.09 pixel RMS in the target region.
The residuals are shown in Figure \ref{fig:Residual-distortion-error}.
Although this represents a large improvement in characterization of
the starfield motion, with better or more complete selection of final
correction functions there is still room for further improvement, as
the residuals still contain visually identifiable patterns. 

\begin{figure}
\begin{centering}
\center{\includegraphics[width=5.5in]{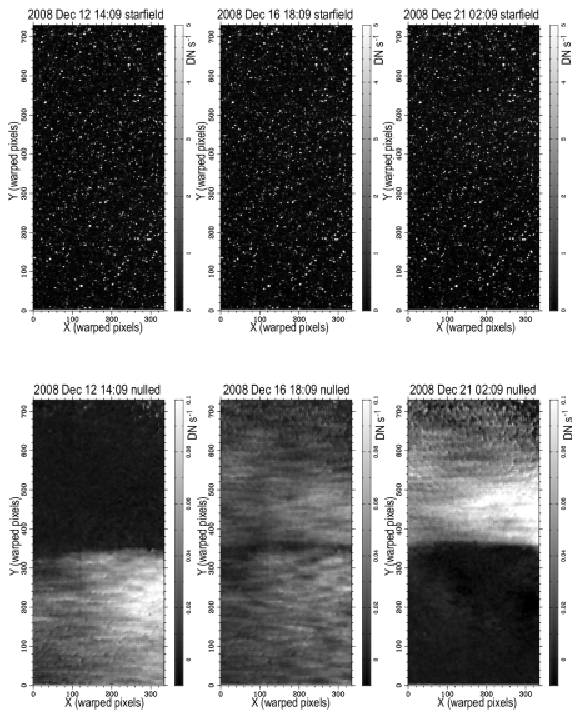}}
\par\end{centering}

\caption{\label{fig:starfield-removal}Freezing in the celestial coordinate
system and starfield subtraction are demonstrated in these three images
from the beginning (left), middle (middle), and end (right) of the dataset. 
Note factor-of-50 difference in the brightness scale between top (frozen) and bottom
(frozen, starfield-subtracted) rows. Solar wind features are faintly
visible but are obscured by residuum from the F coronal subtraction
(see text).}
\end{figure}

The advantage of fitting the pixel displacement function as in Equation
(\ref{eq:distortion}) is that new displacements can be computed using
only the time difference between any two images in the same data set.
We used the fitted $P(x,y)$ and a computed $S(\Delta t)$ to warp
every HI-2A image from 11 Dec 2008 at 00:09:21 through 21 Dec 2008 at 22:09:21
to a central time of 14 Dec 2008 at 00:09:21, to fix the starfield in
the datacube. That fixed starfield was then used to calculate the
median value from every pixel, which formed a background starfield
image that could be subtracted from the cube. Because of the high
gradients around the bright stars, the median subtracted images still
contained paired stellar images due to the subpixel misalignments.
These took the form of dual (positive-going and negative-going) spikes
around each bright star on opposite sides of the PSF core (2-3 pixels
apart) and, near the periphery of the image where the PSF variation
becomes important, triple spikes (two positive-going and one negative-going,
or vice versa) around each bright star due to the changing width of
the PSF. We eliminated these artifacts using a 7$\times$7 pixel running median
filter -- each pixel was replaced by the median of the values in a
7$\times$7 square around itself. The 7$\times$7 median filter effectively blurs
the images by 0.5\degreee, well within the expected motion
blur from 300 km/s solar wind, which is of order 2\degreee\ 
at 0.5 AU distance from the observer. The median-filtered images were then given similar
treatment to the F corona subtraction step above. We identified the
distribution of brightness values in each pixel, selected the 10th
percentile as a background value, and subtracted it from each distorted image.
The results of these steps are given in Figure \ref{fig:starfield-removal}. 

After starfield removal, the remaining visual features have an amplitude
of approximately 0.1 DN $s^{-1}$. Near the center of the data set,
the starfield null is good and essentially no structured stellar artifacts are
seen. Nearer the extremities in either space or time, the starfield
begins to {}``punch through'' due to residual errors in the fitting
function and to PSF variation across the original image field. However,
the background level is no longer dominated by stars but by second-order
effects due to the motion of the F corona during the data collection period.

Even near the center of the image, the starfield remains faintly
visible in the processed data as randomly fluctuating light/dark spots
at the location of the pre-subtraction star.  The fluctuations have no
discernible (to us) spatial or temporal signature that could be
exploited to reduce this noise source further, and they do not have
the characteristic two- or three-peak structure characteristic of
misaligned stars.  We attribute the fluctuations to slight variations
in the effective gain of the detector depending on the phase of the
stellar image relative to the pixel grid as described by
\citet{Jackson2004}. This is the same effect that forced us to use
patch correlation rather than direct centroiding of stars to measure
the distortion function.

Throughout the fitting and analysis process, careful attention to
resampling is critically important because starfields stress the
resampling operation and the usual interpolative shortcuts will spoil
the data.  When the starfield is ``frozen'' by resampling with
bilinear interpolation (as is common in the heliophysics imaging
community), small ripples are visible in the resulting starfield
movie, corresponding to small motions of the stellar centroids of up
to 0.5 pixel in the resampled data and peak amplitude fluctuations of
tens of percent.  In that case the starfield nulling operation is
completely spoiled. To eliminate this type of artifact, we used
spatially variable Jacobian-optimized resampling as described by
\citet{DeForest2004}.

\begin{figure}
\begin{centering}
\center{\includegraphics[width=6in]{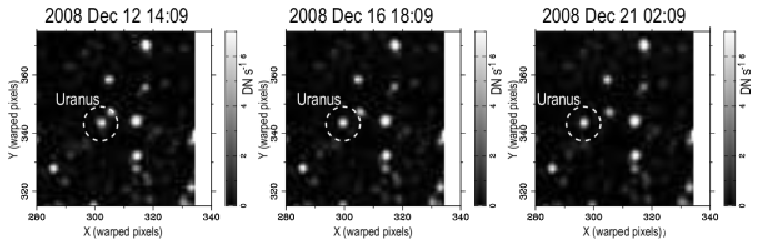}}
\par\end{centering}

\caption{\label{fig:Uranus}Uranus is visible in frozen starfield images from
2008 December. In these three panels, it is the only visible moving
feature (circled and labeled), and has a central brightness 
of approximately 7 DN s$^{-1}$.}
\end{figure}

As an interesting aside, small, faint objects are easily visible in
movies of the frozen starfield. When visually inspecting the frozen-starfield
movie, we noticed a small moving object that proved to be Uranus (Figure
\ref{fig:Uranus}).

\subsection{\label{sub:Residual-F-corona}Residual F corona removal}

After removal of the bulk starfield and fixed F corona, we have
reduced the dynamic range of the artifacts from $\sim$30DN s$^{-1}$ to
$\sim$0.1DN s$^{-1}$, a factor of 300 improvemnt. At this
level, second order effects in the F corona come into play, seen as
the large scale variation in brightness in Figure
\ref{fig:starfield-removal}. The residual brightness enters because
the F corona is not completely time-independent as assumed in
\S\ref{sub:Stationary-background-removal}.  We removed the brightness
variation by making a 3rd order fit to the time varying brightness of
each pixel across the full 11 day observing window, and subtracting
that 3rd order fit from the brightness of each pixel. The automated
fit didn't necessarily find the correct offset to preserve the zero
point, so after subtracting the 3rd order fit, we subtracted from each
pixel the minimum value of the 5$\times$5 pixel median centered on
that pixel, over the whole 12 day run. This process yielded relatively
clean images of the K coronal component of the images, but some
stellar artifacts remained visible.  In particular, although the
spatial median filter in \S\ref{sub:Celestial-background-removal}
removed the large spikes due to stellar images, stars are preserved by
two effects:
\begin{enumerate}
\item the spikes offset the distribution of values in
the remaining pixels in each 5$\times$5 pixel region, so that the median
does change slightly when the distribution of values includes a spike
(though not as much as does the mean); 
\item the instrument point-spread function
is not perfect, so that around each bright star is a small halo of
points that are brighter than deep space by a value comparable to
the signal for which we are looking. 
\end{enumerate}
Both effects give rise to bright artifacts
that are stretched in the direction of stellar motion on the detector,
because they influence the signal in the fixed-focal-plane segment
in \S\ref{sub:Stationary-background-removal}, where the stars are
in motion. These halos are particularly visible in Figure \ref{fig:starfield-removal}
as horizontal bright artifacts. They are difficult to remove with single-image
or pairwise image processing, because they have approximately the
same signal strength and size as features of interest.

\subsection{Fourier moving-object filter}

We removed the remaining celestial-frame fixed artifacts with a Fourier
filtering process. The entire 12 day image set was Fourier transformed
in 3-D to $k_{x},k_{y},\omega$ space, and filtered with a
moving-object filter that rejected all signal associated with wave
speed slower than two image pixels per frame and all fixed signal in
the current (celestial) coordinate system.  The orbital motion of the
starfield in the FOV is approximately 1.2 pixels per frame. For HI-2A,
this motion is in the opposite direction to the solar wind propagation
in the field of view, so our two pixels per frame limit corresponds to
less than one pixel per frame in the original data.  

To perform the filtering, we kept all Fourier energy inside in a cone
with slope of 2 pixels/frame near the $\omega$ axis in the
celestial-frame resampled data, and rejected Fourier energy outside
the cone.  To prevent ringing, we applied a Hanning-style cosine
rolloff with width 1/3 of the base filter width, centered on the
cutoff frequency. This process makes use of the fact that the solar
wind moves in the field of view, while the artifacts are either fixed
or slow-moving, and has been used in a solar context primarily in the
opposite sense - to remove P mode oscillations from solar surface
magnetogram sequences \citep{Hagenaar2005,Lamb2008}.  Fast moving
objects in magnetogram sequences are typically P modes that have high
phase speed, and are rejected, while slow moving objects are the
desired signal.  Here, the sense is reversed, and we preserve moving
solar wind features while rejecting the fixed pattern of the
starfield.

\begin{figure}
\begin{centering}
\center{\includegraphics[width=6in]{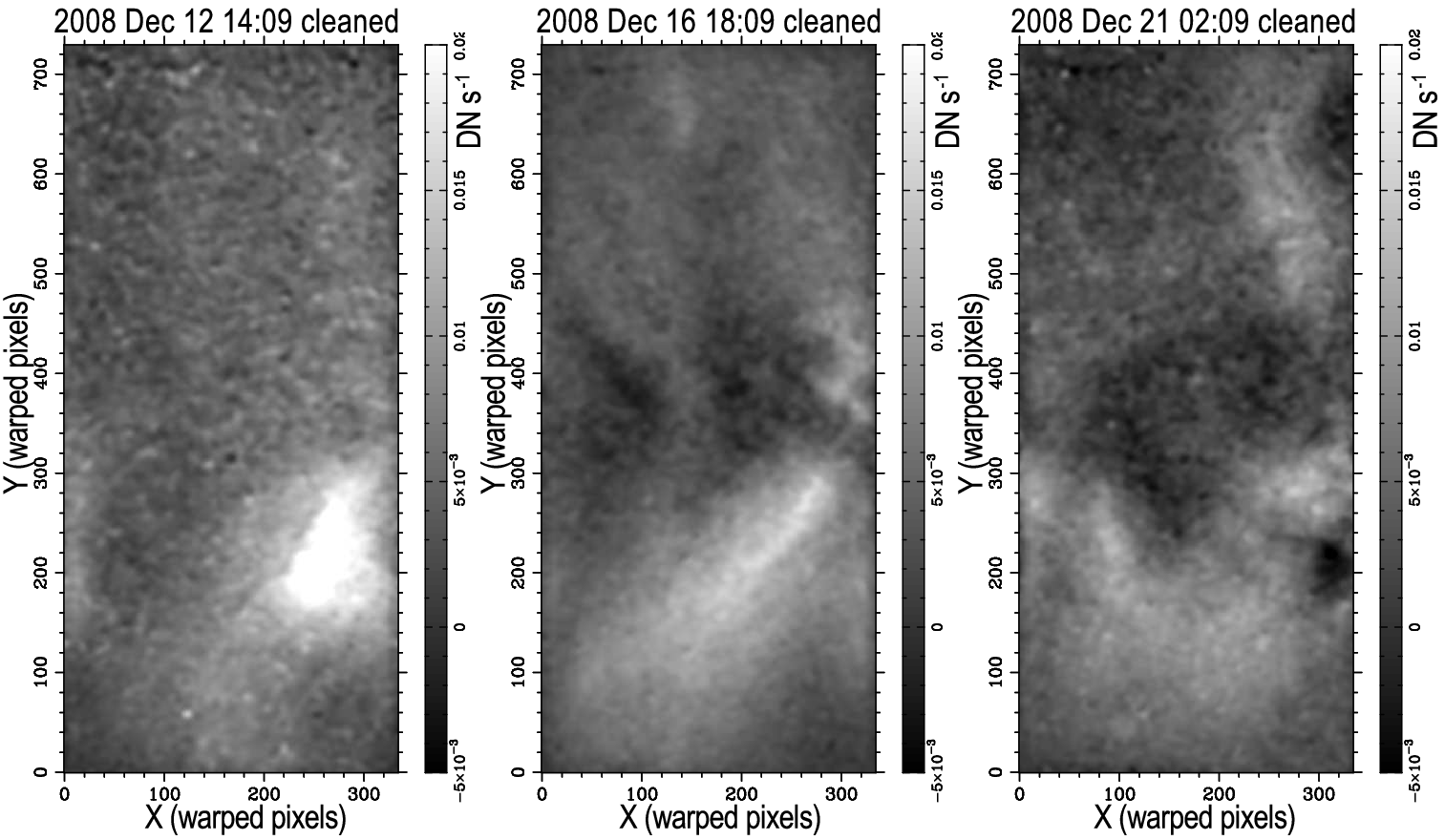}}
\par\end{centering}

\caption{\label{fig:Cleaned-images}Cleaned images, in celestial coordinates,
show solar wind at the 0.02 DN s$^{-1}$ level (full
scale), with discernible features an order of magnitude fainter still.
The brightness scale has been reduced by a factor of 1000
from Figure \ref{fig:raw-data}. The noise level is about $1.5\times10^{-3}$ DN s$^{-1}$
RMS, or about $10^{-4}$ of the raw data amplitude. }

\end{figure}

In addition to slow moving components, we rejected the fixed
($\omega=0$) component of the Fourier image, which eliminates both the
mean brightness and pattern background. This loss was repaired post
facto by identifying the ``zero point'' of each pixel to be the
minimum value of a 25$\times$25$\times$5 voxel average centered on
each voxel in the 3-D $(x,y,t)$ data space.  When the zero point is
subtracted, the remaining signal contains only the brightness component
contained in moving features.  The result is demonstrated in Figure
\ref{fig:Cleaned-images}, which shows the same three frames (out of
121) as in Figure \ref{fig:starfield-removal}, after full processing
in celestial coordinates.

\subsection{Conversion to focal plane coordinates}

\begin{figure}

\center{\includegraphics[width=5.5in]{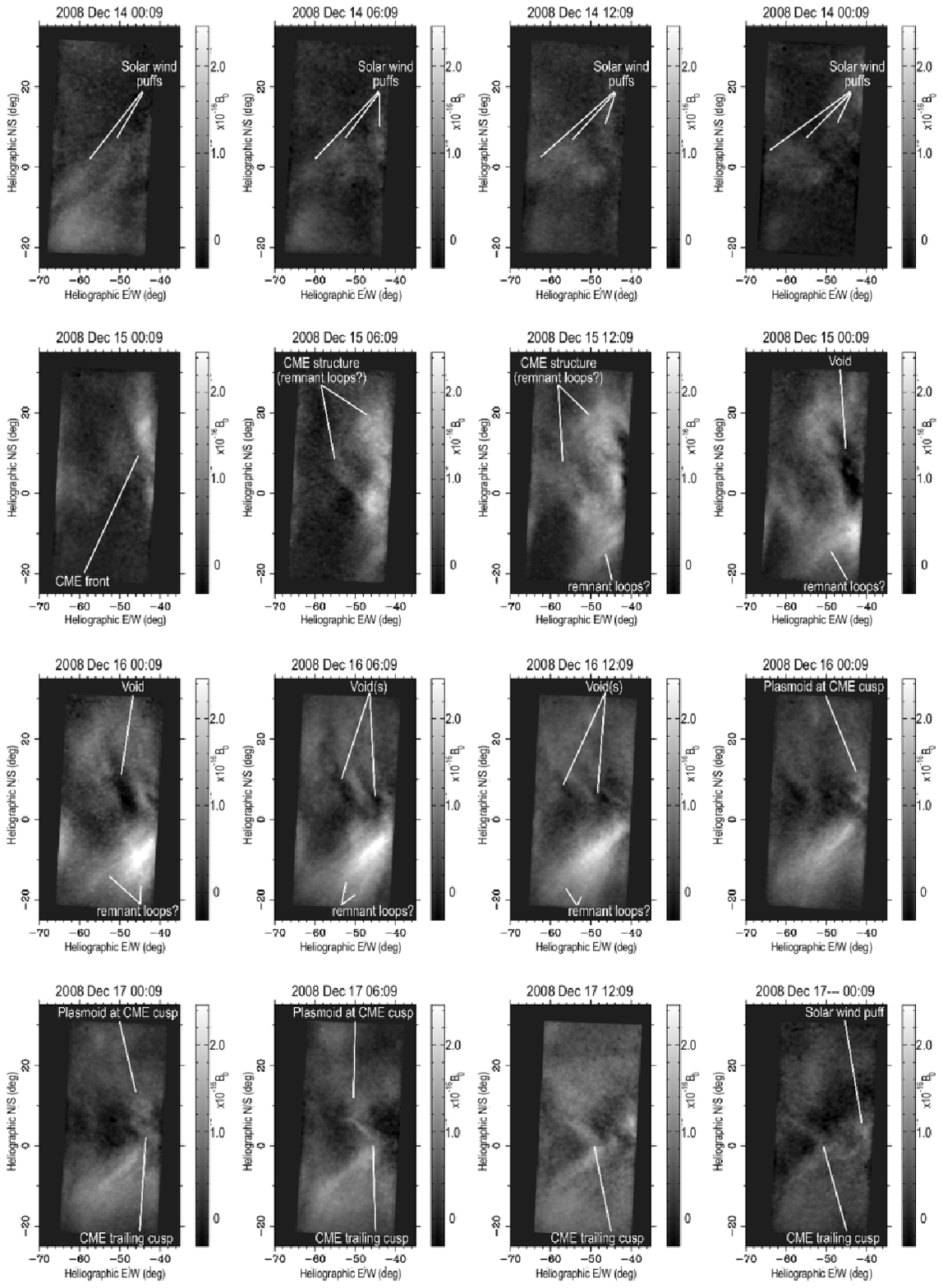}}

\caption{\label{fig:movie-frames}Background-subtracted HI-2 frames, in celestial
coordinates, from 2008 December, show solar wind ``puffs'' and
eddies, a bright CME front, post-front voids, and other related structures.
The images have been sampled at 6 hour cadence and are a subset of
the images in the attached movie file in the digital edition of this article.
In both the images and the movie, the S/N ratio of the brightest features is of
order 30, and faint features down to $3\times 10^{-17}$ \sbright\ are readily seen.}

\end{figure}

After processing, we resampled the filtered images back into focal plane coordinates
to yield a dataset in the original HI-2 coordinate system, with the
background removed. We also applied the preliminary HI-2 calibration
of $1.1\times10^{-14}$ \sbright\ DN$^{-1}$ s (C. Eyles, priv. comm.). 
The final result is shown in Figure
\ref{fig:movie-frames} and, in the digital edition of this article,
in the attached movie file.

\section{Results}

Figure~\ref{fig:movie-frames} shows extracted still frames from the
data set and, in the digital edition of this article, a movie of the
entire data set in original observing coordinates.  The images are
cropped to avoid both Venus and the Earth, and therefore show only a
section of the HI-2 field of view, spanning about
25\degreee$\times$60\degreee.  The FOV moves gradually to the right as time
increases, because the cropping, starfield processing, and Fourier
filtering were carried out in celestial coordinates before being
transformed back into observing coordinates.

The brightness scale of Figure~\ref{fig:movie-frames} is $0-2.5\times
10^{-16}$ \sbright; individual features can be observed at
brightnesses of $\sim 2\times 10^{-17}$ \sbright.  A great deal of
structure appears in the sequence, and we have identified a number of
features that can be isolated from the background solar wind
activity. The level of variance in the solar wind is itself
remarkable. 

HI-2 observations of specific classes of
moving features \citep{Sheeley2008a} have revealed variability in this
altitude range, but to our knowledge this is the first algorithm
capable of fixed computed-background subtraction to reduce image
contamination below the brightness range of fine-scale features in the
solar wind and enabling a direct view of evolving structures as they
propagate.  More detailed scientific analysis will appear in a
following paper; here, we confine ourselves to a first-look
examination of the data.

The CME itself first appears in the image dated 2008-12-15 at 00:09:21
in Figure~\ref{fig:movie-frames} but a number of distinct features in
the solar wind can be identified and tracked prior to its
appearance. They appear across the top row and we have labeled them
``Solar wind puffs''. The second and third row shows the passage of
the CME and a number of accompanying features that can also be
identified and tracked, including ``remnant loops'' that may be
indicative of flux rope structure and that surround an empty cavity or
void.  The structure slightly ahead of the CME in the second and third
frames of the second row (2008-12-15T06:09:21 and T12:09:21) may be
a CME forerunner \citep{Jackson1978} and the bright structure
toward the south of the CME across the third row may be a prominence
or streamer blowout. We have also identified features flowing in the
wake of the CME, two different types have been labeled ``Plasmoid at
CME cusp'' and ``CME trailing cusp'' in the bottom two rows. The
former is aligned near the center of the CME and appears to be either
a related sympathetic eruption or part of the original magnetic
structure comprising the pre-launch coronal field. The latter has a
concave-outward V shape and straddles the solar equatorial region,
meaning that it may be associated with the heliospheric current sheet.

Behind the CME front in Figure~\ref{fig:movie-frames} are two clear,
dark voids.  Voids behind CMEs have been detected remotely via radio
interplanetary scintillation \citep{Tappin1983,Tappin1987,Tappin2010}
but have not been directly imaged before at such high elongations,
because heliospheric imaging studies have either used running
difference images, thereby confusing dark regions with bright regions
that were present in the previous image, or have not achieved the
level of sensitivity required to observe the voids and trailing faint
structure.  We tentatively identify the voids with the cavity in the
the classical three-part CME often observed in coronagraphs
\citep{Sime1984} and also with the region of reduced density often
observed accompanying (and sometimes identified as a signature of)
in-situ magnetic clouds \citep[e.g.][]{Wu2002,Wei2003}. When the feature
impacted the \emph{Wind} spacecraft a void was detected in-situ (see below).

Figure~\ref{fig:J-plot} shows the data sequence in the form of an
elongation-time ``J-plot'', produced by selecting a position angle
slice across an elongation-time-intensity data cube, and utilized
widely by the HI community \citep[e.g.][]{Davies2009}. A number of
position angle slices have been chosen, measured relative to the
ecliptic plane. Many of the features are labeled where they correspond
with those in the movie frames.  

\begin{figure}

\begin{centering}
\center{\includegraphics[width=6in]{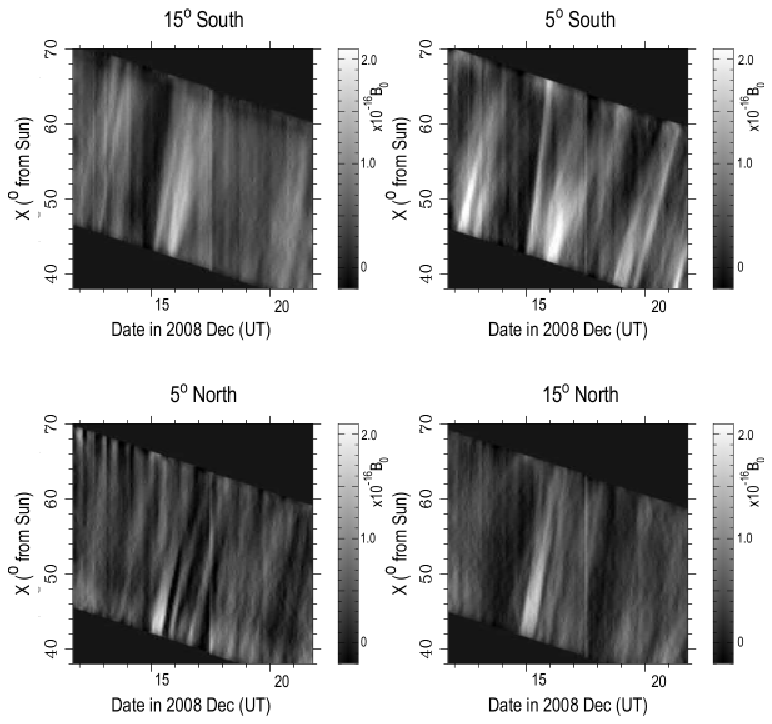}}
\par\end{centering}

\caption{\label{fig:J-plot}Elongation-time ``J-plot'' diagrams for the 11-day
observation period in December 2008 showing multiple features. Each image represents
a different position angle measured from the ecliptic plane: 15\degreee\ south (top left);
5\degreee south (top right); 5\degreee\ north (bottom left); 15\degreee\ north (bottom right).
Only the regions subject to the advanced data processing are shown.}

\end{figure}

\begin{figure}

\center{\includegraphics[width=6in]{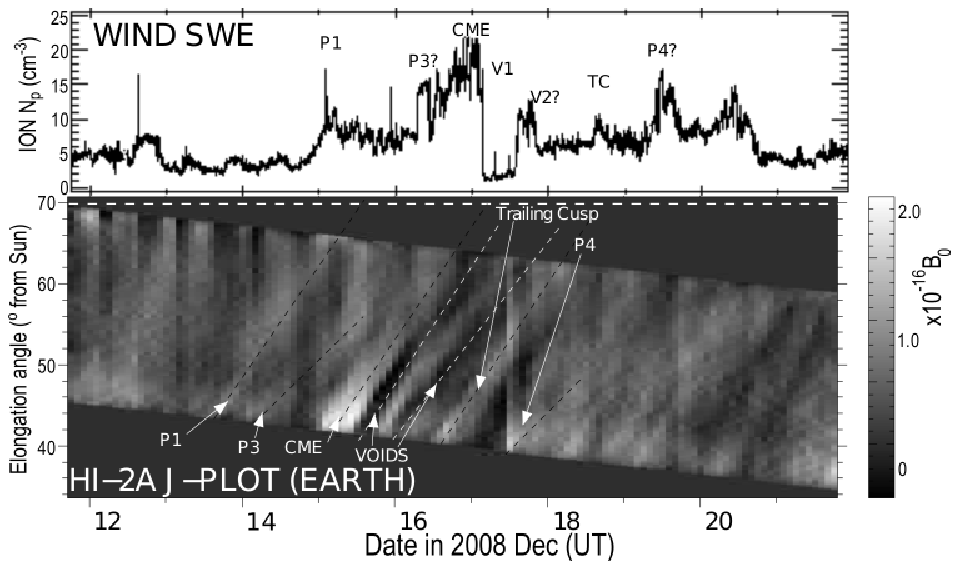}}

\caption{\label{fig:in_situ_comparison}\emph{STEREO}/HI-2A J-plot with
  in-situ density data provided by the \emph{Wind} spacecraft for the
  same time period. These have been produced in the same manner as
  those in Figure~\ref{fig:J-plot}. We have selected the position
  angle to be along the ecliptic, such that the Earth (and Venus) are
  intersected.  The Earth, at 70\degreee\ elongation, is represented
  by a horizontal dashed line.}

\end{figure}

Figure~\ref{fig:in_situ_comparison} shows a J-plot for the ecliptic
plane (+5\degreee\ relative to \emph{STEREO-A} heliographic
coordinates during this time of year) built from the processed data
sequence.  The Earth is indicated as a dashed horizontal line near
70\degreee\ elongation.  This allows direct comparison between the
HI-2 features and in-situ density data, and the density measured by
\emph{Wind}/SWE for the same time period is also shown above.  Most of the
features that we can track in the HI-2 dataset at this heliospheric
latitude also appear in the in-situ data. Several labeled features
from Figure~\ref{fig:movie-frames} are indicated in
Figure~\ref{fig:in_situ_comparison} and extended to the Earth with a
dashed line. A few features (e.g.\ the second puff) do not appear or
cannot be traced all the way to Earth (e.g.\ P4), either because they
miss the Earth's heliospheric longitude are weak against the
background or other brighter features. Three-dimensional analysis is
planned using combined background-subtracted data from both of the
\emph{STEREO} spacecraft and from other auxiliary data sets.

\section{\label{sec:Discussion}Discussion}

We have demonstrated extraction of a quantitative solar wind imaging signal
from a HI-2 image sequence containing 3-4 orders of magnitude brighter
background than the desired data.  We are able to approach, but not achieve,
the photon counting noise limit in the \emph{STEREO}/HI-2 instrument.

The limiting factors in the final image quality of our data set are
(1) stellar background ``print-through'', which we attribute primarily 
to subpixel variations in CCD sensitivity coupled to the relative
phase of the pixel grid and the star's location; and (2) motion blur
from the standard two hour accumulation time for HI-2, with
corresponding blur of up to a few degrees with typical feature speed
ranges \citep[e.g.][]{HowardS2008}. The stellar background print-through
dominates over photon counting noise in the feature size range that
is observable in our data set, by a factor of about 3.

Because we use the image stream itself as a background source,
exploiting differences in behavior and spatial distribution to
distinguish desired data from the background, it is not possible to
extract absolute brightnesses -- only absolute ``excess brightnesses''
of moving features, as is typical for other unpolarized
Thomson-scattering detectors such as \emph{SOHO}/LASCO
\citep{Brueckner1995}. Further, the final filtering step makes use of
the fact that the solar wind apparent speed is fast compared to the
angular orbital speed of the spacecraft, and therefore the processd
data are not suited to imaging of features in the solar wind with
little apparent motion, such as corotating interaction
regions. 

Further improvement of this type of data would require (1) shorter
effective exposure times or better motion compensation; (2) a way to
minimize stellar print-through; and (3) an absolute measurement of the
celestial background independent of the desired Thomson-scattered
signal. Happily, all three are technically feasible: (1) may be
accomplished with a larger instrument or with higher downlinked data
volume; (2) may be accomplished with higher data volume by either
defocusing the optics or dithering the exposure on the CCD between
individual camera exposured during image accumulation; and (3)
polarization can be used to separate the Thomson scattered signal from
the background signal.  Polarization, in particular, is important
because the stars and galaxy, which are far more challenging than the 
F corona to remove, are unpolarized in broadband visible light, and the
slight F coronal polarization follows established models of particulate
scattering \citep{Leinert1997}.

We are able to reveal absolute excess brightness of solar wind and CME
structures with a S/N ratio of up to 30 in 1.5\degreee\ patches of
image, and demonstrate the presence of ``voids'' and complex structure
in the wind stream more than 60\degreee\ (or $\sim$0.85 AU) from the
Sun. Many features that are visible, including ``V'' shaped structures
surrounding the heliospheric current sheet near the plane of the
ecliptic, are to be expected from the wind speed gradient but have not
been observed directly this far from the Sun.

While the presence of structure in the solar wind has been known since
the first in-situ measurements of particle density, and some wind
transients have been detected with the \emph{STEREO}/HI-2 imagers, to our
knowledge this is the first report of quantitative white light imaging
of detailed solar wind structure so far from the Sun.  That is not
surprising given the challenges inherent in separating the bright
starfield from the faint Thomson signal, which requires both deep
subpixel understanding of the instrument's projection function and
careful attention to the mechanics of data resampling.  We were aided
in both of these goals by extensive use of Perl Data Language
\citep{Glazebrook1997}, a scientific computing package that includes a
framework for parameterizing and combining coordinate transformations,
and for optimized data resampling \citep{DeForest2004}.

Efforts currently in progress, to be followed up in future reports,
include three-dimensional reconstruction of feature complexes using
dual spacecraft views and image analysis \citep[e.g.][]{Howard2011},
direct association of CME structure at high elongation with
corresponding structure at the Sun via continuous tracking from launch
to 1 AU; and a more detailed analysis of the small scale features
revealed in the reprocessed data.  

The analysis technique we present here is based on batch processing of
images and is limited to a few days per batch, partly by practical
constraints of computer RAM and partly by motion of the starfield
(which induces a trade-off between length of run and field of view);
however, an open-ended processing pipeline has been prototyped, and
works by merging the results of multiple batch-processed data sets.
That technique and its results will be presented in an article to
follow this one.

\acknowledgements{ }

The authors thank the HI team for the kind use of their calibrated
data. \emph{STEREO}/HI was developed by a collaboration including the
Rutherford Appleton Laboratory and the University of Birmingham (UK),
the Centre Spatial de Li\'{e}ge (Belgium), and the Naval Research
Laboratory (US).  The work for the present study was supported in part
by the NSF/SHINE Competition, Award 0849916, and in part by the NASA
Heliophysics program through grant NNX10AC05G. SJT is supported at NSO
by the USAF under a Memorandum of Agreement. The data reduction
utilized the free Perl Data Language, which is available at
``http://pdl.perl.org''.  The authors also gratefully acknowledge
insightful discussions with S.~Antiochos, S.~Crothers, C.~Eyles,
N.~Schwadron, and N.~Sheeley.

\end{document}